\documentclass[twocolumn,prl,aps,superbib,tightenlines,floatfix]{revtex4}

\usepackage{amsfonts}
\usepackage{amsmath}
\usepackage{amssymb}
\usepackage{graphicx}
\usepackage{bm}

\begin{document}

\title{Ideal switching effect in periodic spin-orbit coupling structures}
\author{S. J. Gong and Z. Q. Yang\cite{ZY}}
\affiliation{Surface Physics Laboratory (National Key Laboratory),
Fudan University, Shanghai, 200433, China.}

\begin{abstract}
An ideal switching effect is discovered in a semiconductor nanowire with a
spatially-periodic Rashba structure. Bistable `ON' and `OFF' states can be
realized by tuning the gate voltage applied on the Rashba regions. The
energy range and position of `OFF' states can be manipulated effectively by
varying the strength of the spin-orbit coupling (SOC) and the unit length of
the periodic structure, respectively. The switching effect of the nanowire
is found to be tolerant of small random fluctuations of SOC strength in the
periodic structure. This ideal switching effect might be applicable in
future spintronic devices.\newline
\end{abstract}

 \maketitle

Spin freedom of electrons in semiconductors can be manipulated efficiently
through the mechanism of spin-orbit couplings (SOCs) \cite{Lutt,Dres,Rash},
which has been confirmed in experiments \cite{Wund}. Among the several types
of SOCs in semiconductors, Rashba SOC \cite{Rash}, which results from
asymmetric electric confinement in nanostructures, is the most attractive
one, due to its strength tuned easily by external gate voltage \cite%
{Nitt,Enge}. Various spintronic devices, such as the spin filter \cite{Koga}%
, spin valve \cite{Mats}, and spin-field-effect transistor \cite{Datt} have
been brought forward in two dimensional electron gases with Rashba
interactions. Since no external magnetic field is required to realize the
control of spin of electrons, all-electrical fabrication of practical
devices has been expected in such kinds of systems \cite{Popescu,Step}.

Very recently, based on Rashba and/or Dresselhaus SOCs, Jiang \textit{et al.
}\cite{Jiang} and Gong \textit{et al}. \cite{Yang} discovered an interesting
switching effect of electronic flow in a one-dimensional electron gas
sandwiched between two electrodes. The transmission coefficient of electrons
in the drain electrode can be varied from 1 to 0 by tuning the SOC strength.
However, in both schemes, the behavior of the switching effect is strongly
dependent on the height of the scattering potentials at the interfaces
between sample and electrodes. With a high interfacial barrier, `ON' state
of the switch can not work effectively: the total transmission peak is too
sharp to gain a stable `ON' state. While with a relatively low barrier,
`OFF' state can not be absolutely reached: there is usually considerable
leakage in the `OFF' state even if the SOC strength is tuned to the maximum
value permitted in current experiments \cite{Nitt,Enge}. And the barrier
height, to our knowledge, can not be controlled effectively by experimental
tacts. All these reduces the feasibility of the practical application of
their switching schemes.

In the present work, an ideal switching effect is found in a one-dimensional
semiconductor quantum wire with spatially-periodic Rashba structure, where
SOC and non-SOC segments connect in series alternately. The principle of the
effect can be rationalized by the transport properties of the electrons in
the wire. When an appropriate magnitude of Rashba strength is provided, an
energy gap can be formed near the boundaries of Brillouin zone due to the
periodic Rashba potential. This causes the incident electrons with energies
in the gap reflected totally. If Rashba strength is tuned to be smaller than
a critical value, all the\ incident electrons can be transmitted. Therefore,
stable `rectangle-type' switching effect can be obtained by controlling the
Rashba SOC. Our further investigation shows that the ideal switching
behavior survives from small fluctuations of the Rashba strengths in the
periodic structure.

\textit{\ }The geometry\textit{\ }we consider is a one-dimensional quantum
wire \cite{Note} with periodic Rashba structure illustrated in Fig.1. Each
periodic unit consists of one non-SOC segment and one SOC segment with the
same length of $a/2$ ($a$ is set at 24 nm in the following calculations
except the case in Fig. 2 (b)). The symbol of $V_{g}$ in the figure
expresses the applied gate voltage to control the Rashba strength. Note that
in our model the SOC and non-SOC segments are composed of the same
semiconductor material. Therefore, if gate voltage is removed (i.e. Rashba
SOC is neglected), all the segments unite into a homogeneous structure. In
the calculation, an electron wave is injected from the left to the right
along $x$ direction.

\begin{figure}[tbph]
\begin{center}
\resizebox{8cm}{!}{\includegraphics*[56,363][546,579]{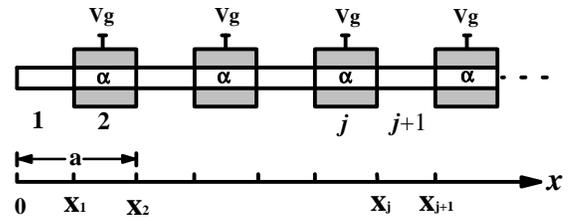}}
\end{center}
\caption{Schematic diagram of the switch device: a one-dimensional
quantum wire along $x$ direction with periodic Rashba structure. }
\end{figure}

The Hamiltonian in SOC segment can be written as:%
\begin{equation}
H=\frac{p_{x}^{2}}{2m^{\ast }}-\dfrac{\alpha }{\hbar }\sigma _{y}p_{x},
\end{equation}%
where the effective mass of electrons $m^{\ast }$ is set as 0.067 $m_{e\text{
}}(m_{e}$ is the mass of the free electron), $\sigma _{y}$ is the vector of
the Pauli matrix, and $p_{x}$ is the $x$-component of the momentum operator.
The parameter $\alpha $ describes the SOC strength. To determine the final
transmission coefficient after propagating through the whole quantum wire,
we need consider the transmission process of an electron with energy $E$
through one unit, and obtain the transfer matrix. To provide a clear
illustration, we lable the segments in series: $1,$ $2$ $...j,$ $j+1...$, as
shown in Fig.1. The even $j$ stands for SOC segments, and the wave function
in it can be expressed as: $\psi _{j}=a_{j}e^{ik^{+}x}\left\vert \uparrow
\right\rangle +b_{j}e^{-ik^{-}x}\left\vert \uparrow \right\rangle
+c_{j}e^{ik^{-}x}\left\vert \downarrow \right\rangle
+d_{j}e^{-ik^{+}x}\left\vert \downarrow \right\rangle $, where$\qquad
k^{+}=\left( \alpha +\sqrt{\alpha ^{2}+\frac{2\hbar ^{2}E}{m^{\ast }}}%
\right) \frac{m^{\ast }}{\hbar ^{2}}$, $k^{-}=\left( -\alpha +\sqrt{\alpha
^{2}+\frac{2\hbar ^{2}E}{m^{\ast }}}\right) \frac{m^{\ast }}{\hbar ^{2}}$.
The denotions $\left\vert \uparrow \right\rangle $ and $\left\vert
\downarrow \right\rangle $ express the eigenspinor states $\binom{1}{i}$ and
$\binom{1}{-i}$, respectively. Similarly, odd $j$ stands for the non-SOC
segments, where the wave function can be written in the same form as SOC
segment with, however, different wave vectors: $k^{^{+}}=k^{-}=k_{0}$ ($%
k_{0}=\sqrt{\frac{2m^{\ast }E}{\hbar ^{2}}}$).

Using boundary conditions at the interfaces of non-SOC/SOC, i.e. the
continuous conditions of wave functions and conservation ones of the current
\cite{Mats,Zuli, Sun}, we can get the following transfer matrix for the wave
functions at $j=2$ and $1$ segments.%
\begin{equation}
\left(
\begin{array}{c}
a_{2} \\
b_{2} \\
c_{2} \\
d_{2}%
\end{array}%
\right) =L_{2}^{-1}\left( x_{1}\right) q_{2}^{-1}q_{1}L_{1}\left(
x_{1}\right) \left(
\begin{array}{c}
a_{1} \\
b_{1} \\
c_{1} \\
d_{1}%
\end{array}%
\right) ,
\end{equation}

where

$L_{1}\left( x_{1}\right) =\left(
\begin{array}{cccc}
{\Large e}^{ik^{+}x_{1}} & 0 & 0 & 0 \\
0 & {\Large e}^{-ik^{-}x_{1}} & 0 & 0 \\
0 & 0 & {\Large e}^{ik^{-}x_{1}} & 0 \\
0 & 0 & 0 & {\Large e}^{-ik^{^{+}}x_{1}}%
\end{array}%
\right) ,$

$q_{1}=\left(
\begin{array}{cccc}
1 & 1 & 0 & 0 \\
0 & 0 & 1 & 1 \\
\frac{\hbar k^{+}}{m^{\ast }} & -\frac{\hbar k^{-}}{m^{\ast }} & 0 & 0 \\
0 & 0 & \frac{\hbar k^{-}}{m^{\ast }} & -\frac{\hbar k^{+}}{m^{\ast }}%
\end{array}%
\right) ,$

$q_{2}=\left(
\begin{array}{cccc}
1 & 1 & 0 & 0 \\
0 & 0 & 1 & 1 \\
\frac{\hbar k^{0}}{m^{\ast }}-\frac{\alpha }{\hbar } & -\frac{\hbar k^{0}}{%
m^{\ast }}-\frac{\alpha }{\hbar } & 0 & 0 \\
0 & 0 & \frac{\hbar k^{0}}{m^{\ast }}+\frac{\alpha }{\hbar } & -\frac{\hbar
k^{0}}{m^{\ast }}+\frac{\alpha }{\hbar }%
\end{array}%
\right) ,$

$L_{2}\left( x_{1}\right) =\left(
\begin{array}{cccc}
{\Large e}^{ik^{0}x_{1}} & 0 & 0 & 0 \\
0 & {\Large e}^{-ik^{0}x_{1}} & 0 & 0 \\
0 & 0 & {\Large e}^{ik^{0}x_{1}} & 0 \\
0 & 0 & 0 & {\Large e}^{-ik^{0}x_{1}}%
\end{array}%
\right) .$

Note that $L_{1}\left( x_{1}\right) $ and $L_{2}\left( x_{1}\right) $ are
related with the coordinates, while $q_{1}$ and $q_{2}$ are not. The
transfer matrix for the wave function in the $j$th segment can be deduced
as: $M_{j}=L_{j}^{-1}\left( x_{j-1}\right) q_{j}^{-1}q_{j-1}L_{j-1}\left(
x_{j-1}\right) \cdot \cdot \cdot L_{3}^{-1}\left( x_{2}\right)
q_{3}^{-1}q_{2}L_{2}\left( x_{2}\right) \cdot L_{2}^{-1}\left( x_{1}\right)
q_{2}^{-1}q_{1}L_{1}\left( x_{1}\right) $, from which the coefficients $%
\left(
\begin{array}{c}
a_{j} \\
b_{j} \\
c_{j} \\
d_{j}%
\end{array}%
\right) $ in the $j$th segment can be expressed as:

\begin{equation}
\left(
\begin{array}{c}
a_{j} \\
b_{j} \\
c_{j} \\
d_{j}%
\end{array}%
\right) =M_{j}\left(
\begin{array}{c}
a_{1} \\
b_{1} \\
c_{1} \\
d_{1}%
\end{array}%
\right) .
\end{equation}

From Eq. (3), the transmitted wave function in the $j$th segment can be
obtained if the incident wave function is known. The total transmission
coefficient ($T$) of spin up and down states in the $j$th segment is then
calculated. In the switch scheme, `$T=1$' and `$T=0$' correspond to ideal
`ON' and `OFF' states of the outgoing wave, respectively.

In order to understand the switching effect well, we calculate the band
structure of the periodic structure by using the plane wave method. The wave
function can be expressed as: $\psi =\psi _{1k}(x)\left(
\begin{array}{c}
1 \\
0%
\end{array}%
\right) +\psi _{2k}(x)\left(
\begin{array}{c}
0 \\
1%
\end{array}%
\right) $, where $\psi _{1k}(x)$ and $\psi _{2k}(x)$ are expanded in plane
waves: $\psi _{1k}(x)=\frac{1}{\sqrt{L}}\sum%
\limits_{K_{n}}C_{1k}(K_{n})e^{i(K_{n}+k)x},$ $\psi _{2k}(x)=\frac{1}{\sqrt{L%
}}\sum\limits_{K_{n}}C_{2k}(K_{n})e^{i(K_{n}+k)x}$, where $L$ is the total
length of the nanowire, $K_{n}$ is the reciprocal lattice vector, $K_{n}=n%
\frac{2\pi }{a}$. The Rashba interaction is modulated periodically along $x$
direction (see Fig.1), and expanded as $\alpha (K_{n})=\frac{1}{a}%
\int_{x_{1}}^{x_{2}}\alpha e^{-iK_{n}x}dx.$ Solving the schr\"{o}dinger
equation in reciprocal space, we get two coupling equations:

\begin{equation}
\begin{array}{l}
\sum\limits_{K_{n}}(\frac{(K_{n}+k)^{2}}{2m}-E_{k})C_{1k}(K_{n})\delta
(K_{n}-K_{m})- \\
\frac{i}{2}\sum\limits_{K_{n}}(K_{m}+K_{n}+2k)\alpha
(K_{m}-K_{n})C_{2k}(K_{n})=0,%
\end{array}%
\end{equation}

\begin{equation}
\begin{array}{l}
\sum\limits_{K_{n}}(\frac{(K_{n}+k)^{2}}{2m}-E_{k})C_{2k}(K_{n})\delta
(K_{n}-K_{m})+ \\
\frac{i}{2}\sum\limits_{K_{n}}(K_{m}+K_{n}+2k)\alpha
(K_{m}-K_{n})C_{1k}(K_{n})=0.%
\end{array}%
\end{equation}%
For each $K_{m}$, there are two equations like above. If the total number of
plane waves used is $N$, there will be $2N$ coupling equations,
corresponding to $2N$ coefficients $\{C_{1k}(K_{m}),C_{2k}(K_{m})$, $m=1,N\}$%
. The eigenvalues at each $K_{m}$ point can be solved by diagonalizing the
secular equation.

The transmission coefficient as a function of the incident energy of
electrons at different Rashba strengths are shown in Fig.2(a). The striking
feature$\ $in the figure is the appearance of energy gaps, within which the
transmission coefficient $T=0$. When Rashba strength $\alpha =0.03$ a.u. ($1$
a.u.$=1.44\times 10^{-9}$ eVm), the width of the gap is about $0.5$ meV
(roughly from 9.1 to 9.6 meV). That means the incident electrons with the
energies in this range will be reflected totally by the periodic structure.
The width of the gap is found to be sensitively dependent on the Rashba
strength. It is clear that the gap becomes larger with the increase of the
Rashba strength. In addition, the gap is also related with the number of the
repeated periodic units in the structure (the repeated number is set at $100$
in the calculation for the periodic structure). At the same Rashba strength,
the gap width will increase with the increase of the number of the periodic
units till it reaches a saturated value. A larger number of units and a
stronger $\alpha $ are expected to produce a wider gap. In the practical
case, appropriate $\alpha $ value and number of units may be chosen.

\begin{figure}[tbph]
\begin{center}
\resizebox{8cm}{!}{\includegraphics*[320,516][499,669]{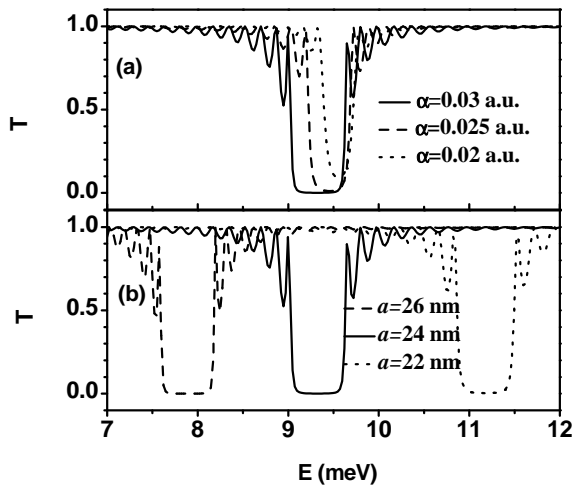}}
\end{center}
\caption{(a) Transmission coefficient as a function of incident
energy of electrons at different Rashba strengths; (b) Transmission
coefficient as a function of incident energy with different lengths
of unit cell. The Rashba strength is taken as $\alpha =0.03$ a.u..}
\end{figure}

To realize the switching effect, we hope that the Fermi energy of the
incident electrons is located within the energy gap, so that electrons can
not transmit through the quantum wire. This state then corresponds to `OFF'\
state of a switch. In Fig.2(b), we fix the SOC strength $\alpha =0.03$ a.u.$%
, $ and pay our attention to the gap\ position under different lengths of
unit cell. It can be seen that the gap shifts toward lower energy region
with the increase of $a$, which can be rationalized by the property of band
structure (in the following). Therefore, by selecting appropriate lengths of
$a$, the position of the energy gap can be modulated according to the
position of the Fermi energy.

From Fig.2(a), we also find that only when the incident energy of the
electrons is within the energy gap, Rashba SOC has the decisive contribution
to the transmission coefficient. Beyond the gap, the contribution of the
Rashba SOC is negligible ($T\simeq 1$). The oscillations of transmission
varying from $T=1$ to $T=0$ as the energy approaches to the position of the
energy gap can be ascribed to the abrupt transition from SOC/non-SOC
interface \cite{Reyn}. To clearly illustrate the contribution of the Rashba
SOC, we plot the dependence of transmission coefficient on the Rashba
strength in Fig.3, in which the incident energies are given as $9.2,$ $9.3,$
and $9.4$ meV. All of the energies are located in the transmission gap shown
by the case of solid curve in Fig.2(a). Obviously, we obtain a binary
\textquotedblleft rectangle-type" transmission behavior with values of $1$
and $0$ by tuning the Rashba strength continuously. For a given energy (for
example, $E=9.3$ meV), we can find a critical value $\alpha _{c}$ ($\alpha
_{c}$ corresponds to the peak in the solid curve). When $\alpha <\alpha _{c}$%
, a nearly total transmission is achieved, corresponding to `ON'\ state.
When $\alpha >\alpha _{c},$ no electron can be transmitted, corresponding to
`OFF' state. It is found that small incident energy corresponds to large $%
\alpha _{c}$, which can be illustrated by the trends of $T$ as a function of
$E$ at different Rashba strengths shown in Fig.2(a). In reality, the
incident energies of the electrons may range from $E_{1}$ to $E_{2}$
(suppose $E_{1}<E_{2}),$ we need to find $\alpha _{c1}$ (corresponding to $%
E_{1})$ and $\alpha _{c2}$ $($corresponding to $E_{2}).$ When $\alpha
>\alpha _{c1},$ the switch is `OFF', when $\alpha <$ $\alpha _{c2},$ the
switch is `ON'.

\begin{figure}[tbph]
\begin{center}
\resizebox{8cm}{!}{\includegraphics*[150,447][446,665]{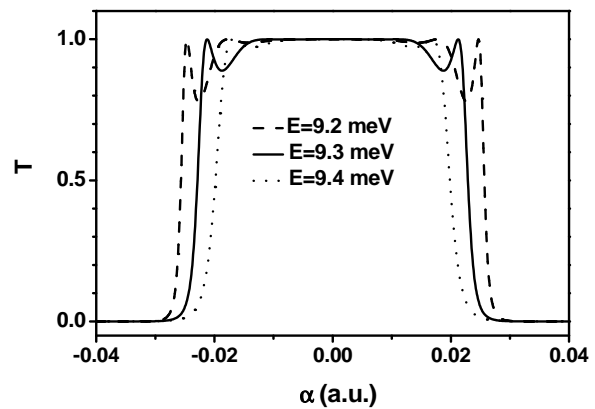}}
\end{center}
\caption{Transmission coefficient as a function of the Rashba
strength with different incident energies.}
\end{figure}

To gain a deep insight into the properties of the spintronic switch, we
investigate the band structure of the one-dimensional system with periodic
Rashba potential. Figure 4(a) shows the band structures without ($\alpha =0$%
) and with Rashba SOC ($\alpha =0.03$ a.u.), respectively. Comparing the
solid curve and the dotted one, we find that due to the Rashba spin-orbit
interaction, the degenerate band structure splits into two subbands: one is
for spin-up and the other for spin-down. Here we emphasize the energy gap
near the boundaries of Brillouin Zone. With the same parameters as in the
case of \ solid curve in Fig.2(a), the gap width in Fig.4(a) is also about
0.5 meV from 9.1 to 9.6 meV. There is, in fact, difference in the geometries
between Fig.4(a) and Fig.2(a). For the band structure calculation, the
one-dimensional system is infinitely long, while the spintronic switch is a
quantum wire with finite length. The fact that Fig.4(a) and Fig.2(a) produce
almost the same gap demonstrates that the length of the periodic Rashba
structure in Fig.2(a) is long enough to be equivalent to the infinite one.
With weak Rashba strength, the difference from the two structures can be
observed. In Fig.2(a), when Rashba strength is decreased to 0.02 a.u., the
energy gap is smeared out. If we increase the number of the periodic unit
cell, the gap will be opened.

\begin{figure}[tbph]
\begin{center}
\resizebox{8cm}{!}{\includegraphics*[139,408][464,645]{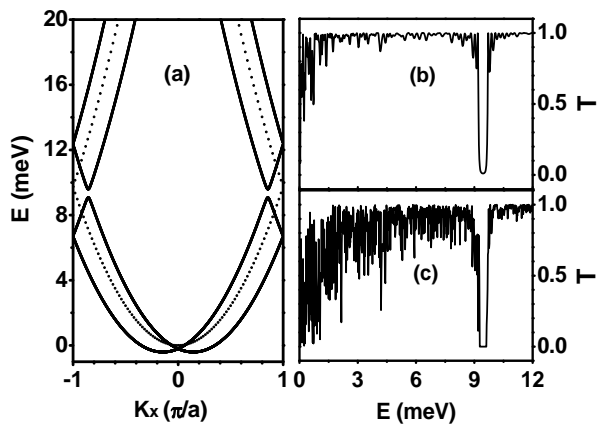}}
\end{center}
\caption{(a) Band structures of one-dimensional quantum wires.
Dotted curve: no Rashba interaction, solid curve: Rashba strength
$\alpha =0.03$ a.u.; (b) Transmission coefficient versus the
incident energy in disordered structure. The number of units in the
structure is 100; (c) The same as in (b) with, however, the number
of units being 1000.}
\end{figure}

In a practical case, we usually can not get a perfect periodic structure,
for example, $\alpha _{j}$ may fluctuate from its set value. Here we
consider a disordered SOC structure, i.e. the Rashba strengths in SOC
segments are randomly given, and investigate its switching effect. Figure
4(b) is the case that Rashba strengths randomly fluctuate from 0.02 a.u. to
0.03 a.u.. Compared Fig.4(b) with the dashed curve of Fig.2(a), whose Rashba
strength is set at 0.025 a.u. (the average value of 0.02 a.u. and 0.03
a.u.), it is found that the energy gaps in the two cases show little
difference. Therefore, it can be inferred that the switch effect we obtained
is tolerant of such disorder. In additional, to our knowledge, in a
one-dimensional system, the presence of disorder will induce localized
states of electrons, which is a critical difference between periodic and
disordered systems. Therefore, the `OFF' state may be achieved due to the
localized states in an ideal disordered system. For example, if the number
of periodic units in the structure increases from 100 in Fig.4(b) to 1000 in
Fig.4(c), the small dips in the energy region of 2.0 to 8.0 meV will become
deeper. Some gaps may form with the further increase of the length of the
structure.

\textit{Conclusion: }A perfect switching effect of electronic flow is found
in a one-dimensiaonl nanowire with spatially-periodic Rashba spin-orbit
coupling. Stable `rectangle-type' switching effect is obtained by
controlling the Rashba SOC strength. The switch effect behaves fairly well
even if the fluctuations of Rashba strengths destroy the periodic structure
to some extent.

The authors are grateful to Prof. R. B. Tao at Fudan University for very
helpful discussion. This work was supported by\textbf{\ }the National
Natural Science Foundation of China with grant Nos.10304002 and 1067027, the
Grand Foundation of Shanghai Science and Technology (05DJ14003), PCSIRT.


\begin{thebibliography}{*}
\bibitem{Note} For example, the model can describe well a quasi-1DEG system
with width less than 20 nm, since the first confinement energy of the
geometry is found to be higer than 20 meV and the Fermi energy of a
semiconductor 2DEG is generally about or below 20 meV.
\end{thebibliography}
\end{document}